\documentclass[prl,12pt]{revtex4}
\pdfoutput=1 
\usepackage{amsmath}
\usepackage{graphicx}
\usepackage{subfigure}
\usepackage{bm}

\def\d{\mathrm{d}}

\begin{document}
\hspace{11.5cm}{MAN/HEP/2012/12}

\title{An AdS/QCD holographic wavefunction for  the ${\boldsymbol{\rho}}$ meson and diffractive ${\boldsymbol{\rho}}$ meson electroproduction}

\author{J. R. Forshaw}
\affiliation{Consortium for Fundamental Physics, School of Physics \& Astronomy, University of Manchester,
Oxford Road, Manchester M13 9PL, U.K.}
\email{jeff.forshaw@manchester.ac.uk} 
\author{R. Sandapen}
\affiliation{D\'epartement de Physique et d'Astronomie, Universit\'e de Moncton,
Moncton, N-B. E1A 3E9, Canada.}
\email{ruben.sandapen@umoncton.ca} 

\begin{abstract}
We show that AdS/QCD generates predictions for the rate of diffractive $\rho$-meson
electroproduction that are in agreement with data collected
at the HERA electron-proton collider. 
\end{abstract}
\keywords{}
\maketitle

\section{Introduction}
To date, the correspondence between string theory in five-dimensional
anti-de Sitter (AdS) space and four-dimensional QCD has enjoyed a
number of successes (see
\cite{Erdmenger:2007cm,CasalderreySolana:2011us,Costa:2012fw,deTeramond:2012rt} and
references therein).
In this letter, we demonstrate another
success by showing that parameter-free AdS/QCD wavefunctions for the $\rho$
meson lead to predictions for the rate of diffractive $\rho$ meson
production ($\gamma^* p \to \rho p$) that agree with
the data collected at the HERA $ep$ collider.

In previous papers \cite{Forshaw:2010py,Forshaw:2011yj}, we took a
phenomenological approach and extracted the light-front wavefunctions
of the $\rho$ meson using the HERA data. We follow the same formalism
here, except that we now use the AdS/QCD wavefunctions predicted in \cite{Brodsky:2007hb,Brodsky:2008pg}.

\section{The AdS/QCD wavefunction}
Brodsky and de T\'eramond have recently shown that, in what they call a
first semiclassical approximation to light-front QCD \cite{deTeramond:2008ht}, the meson
wavefunction can be written in the
following factorized form 
\begin{equation}
\phi(x,\zeta, \varphi)=\frac{\Phi(\zeta)}{\sqrt{2\pi \zeta}} f(x) \mathrm{e}^{i L \varphi} 
\label{factorized-lc}
\end{equation}
where $L$ is the orbital quantum number and $\zeta=\sqrt{x(1-x)} b$ ($x$
is the light-front longitudinal momentum fraction of the quark
and $b$ the quark-antiquark transverse separation). The function $\Phi(\zeta)$ satisfies a
Schr\"odinger-like wave equation
\begin{equation}
\left(-\frac{\d^2}{\d \zeta^2} - \frac{1-4L^2}{4 \zeta^2} + U(\zeta) \right) \Phi (\zeta)=M^2 \Phi (\zeta) \;,
\label{LFeigenvalue}
\end{equation}
where $U(\zeta)$ is the confining potential defined at equal light-front time. After identifying $\zeta$ with
the co-ordinate in the fifth dimension, Eq.~\eqref{LFeigenvalue}
describes the propagation of spin-$J$ string modes, in which
case $U(\zeta)$ is determined by the choice for the dilaton
field. We shall use the soft-wall
model~\cite{Karch:2006pv}, in which 
\begin{equation}
 U(\zeta)=\kappa^4 \zeta^2 + 2\kappa^2(J-1) \;.
\label{quadratic-dilaton}
\end{equation}
This potential encodes the 
confinement dynamics of QCD and the challenge remains to derive it from 
first-principles QCD.

Solving Eq.~(\ref{LFeigenvalue}) with this potential results in eigenvalues
\begin{equation}
 M^2=4\kappa^2(n+J/2+L/2) \; , 
\label{mass-spectrum}
\end{equation}
which reproduces the correct meson mass spectrum. In particular, it
predicts a massless pion ($S=0, n=0, L=0$) and $M_\rho^2=2\kappa^2$ for the $\rho$ meson
($S=1, n=0, L=0$).  The corresponding
eigenfunctions are \cite{Vega:2009zb}
 \begin{equation}
 \Phi (\zeta)= \kappa \sqrt{2\zeta} \exp \left(-\frac{\kappa^2 \zeta^2}{2}\right)  \;.
\end{equation}
It remains to
specify the function $f(x)$ in Eq.~\eqref{factorized-lc}. This can be done by
comparing the expressions for the pion EM form factor
obtained in the light-front formalism and in AdS
space \cite{Brodsky:2007hb}  and it results in 
\begin{equation}
f(x) = {\cal{N}}\sqrt{x(1-x)} \;. 
\end{equation}
The resulting wavefunction is thus 
\begin{equation}
 \phi(x,\zeta)= \mathcal{N} \frac{\kappa}{\sqrt{\pi}}\sqrt{x(1-x)} \exp \left(-\frac{\kappa^2 \zeta^2}{2}\right)~,
\label{lcwf-massless-quarks}
\end{equation}
where $\mathcal{N}$ is a normalisation constant. Assuming the meson is
dominated by its leading $q\bar{q}$ Fock component, $\mathcal{N}$ is fixed by
\begin{equation}
\int {\d}^2{\mathbf{b}} \; \d x \; |\phi(x,\zeta)|^2 = \int_0^1
\frac{\d x}{x(1-x)} f^2(x) = 1~. \label{eq:norm0}
\end{equation}
 Brodsky and de
T\'eramond also have a prescription to account for non-zero quark
masses \cite{Brodsky:2008pg}: A Fourier transform to $k$-space gives 
\begin{equation}
\tilde{\phi}(x,k) \propto \frac{1}{\sqrt{x(1-x)}} \exp \left(-\frac{M^2_{q\bar{q}}}{2\kappa^2} \right) ~,
\end{equation}
where the invariant mass squared of the $q\bar{q}$ pair is
$M^2_{q\bar{q}}=k^2/(x(1-x))$. For massive quarks, the invariant mass should rather be
$M^2_{q\bar{q}}=(k^2 + m_f^2 )(x(1-x))$. After substituting this into
the wavefunction and Fourier
transforming back to transverse position space, one obtains the final
form of the AdS/QCD wavefunction:
\begin{equation}
 \phi(x,\zeta)= N \frac{\kappa}{\sqrt{\pi}}\sqrt{x(1-x)} \exp \left(-\frac{\kappa^2 \zeta^2}{2}\right) \exp\left(-\frac{m_f^2}{2\kappa^2 x (1-x)} \right)~,
\label{lcwf-massive-quarks}
\end{equation}
and $N$ is fixed by the generalization of Eq.~(\ref{eq:norm0}):
\begin{equation}
\int_0^1 \frac{\d x}{x(1-x)}
f^2(x)\exp\left(-\frac{m_f^2}{\kappa^2x(1-x)} \right) = 1~. \label{eq:norm}
\end{equation}
This is rather similar to the original Boosted Gaussian (BG)
wavefunction discussed in \cite{Nemchik:1996cw,Forshaw:2003ki}
\begin{equation}
\phi^{{\mathrm{BG}}} (x,\zeta) \propto x(1-x) \;
\exp \left(\frac{m_f^{2}R^{2}}{2}\right)
\exp \left(-\frac{m_f^{2}R^{2}}{8 x(1-x)}\right) \; \exp \left(-\frac{2 \zeta^{2}}{{R}^{2}}\right) \;.
\label{original-boosted-gaussian} 
\end{equation}
If $R^2=4/\kappa^2$ then
the two wavefunctions differ only by a factor of $\sqrt{x(1-x)}$,
which is not surprising given that
in both cases confinement is modelled by a harmonic oscillator.
In what follows we shall consider a parameterization that accommodates
both the AdS/QCD and the BG wavefunctions: 
\begin{equation}
 \phi(x,\zeta) \propto [x(1-x)]^\beta \exp \left(-\frac{\kappa^2 \zeta^2}{2}\right) \exp\left(-\frac{m_f^2}{2\kappa^2 x (1-x)} \right)~.
\label{lcwf-massive-quarks-fit}
\end{equation}

\section{Comparing to data, QCD Sum Rules and the lattice}

To compute the cross-section for $\gamma^* p \to \rho p$ we use the
dipole model of high-energy scattering
\cite{Nikolaev:1990ja,Nikolaev:1991et,Mueller:1993rr,Mueller:1994jq}. In
this approach, the scattering amplitude is a convolution of the photon
and vector meson $q\bar{q}$ wavefunctions with the total cross-section
to scatter a $q\bar{q}$ dipole off a proton. QED is used to determine
the photon wavefunction and the dipole cross-section can be extracted
from the precise data on the deep-inelastic structure
function $F_2$. The details of this procedure can be found in
\cite{Forshaw:2011yj,Forshaw:2003ki}. All that remains is to specify the wavefunction of the meson.

The meson's light-front wavefunctions can be written in terms of the
AdS/QCD wavefunction $\phi(x,\zeta)$ \cite{Forshaw:2011yj}. For longitudinally
polarized mesons:
\begin{equation}
\Psi^{L}_{h,\bar{h}}(b,x) = \frac{1}{2\sqrt{2}}
\delta_{h,-\bar{h}} 
\left( 1 +  \frac{ m_{f}^{2} -  \nabla^{2}}{M_{\rho}^2\; x(1-x)}\right) \phi_L(x,\zeta) ~,
\label{nnpz_L}
\end{equation}
where $\nabla^2 \equiv \frac{1}{b} \partial_b + \partial^2_b$ and
$h$ ($\bar{h}$) are the helicities of the quark (anti-quark). 
For transversely polarized mesons:
\begin{equation}
\Psi^{T=\pm}_{h,\bar{h}}(b, x) = \pm [i e^{\pm i\theta} 
( x \delta_{h\pm,\bar{h}\mp} - (1-x) \delta_{h\mp,\bar{h}\pm}) 
\partial_{b}+ m_{f}\delta_{h\pm,\bar{h}\pm}] \frac{\phi_T(x,\zeta)}{2x(1-x)}~,
\label{nnpz_T}
\end{equation}
where $be^{i\theta}$ is the complex form of the transverse separation,
$\mathbf{b}$. Rather than using Eq.~(\ref{eq:norm}) to fix the
normalization of $\phi$ we impose 
\begin{equation}
\sum_{h,\bar{h}}\int \d^{2}{\mathbf{b}} \, \d x  \,
|\Psi^{\lambda}_{h,\bar{h}}(b, x)|^{2} = 1 ~,
\label{normalisationTL}
\end{equation}
where $\lambda=L,T$. This means we allow for a polarization dependent
normalization (hence the subscripts on $\phi_{L,T}$). For longitudinal
polarization, the difference between the two normalization
prescriptions leads only to sub-leading
differences $\sim M_\rho^2/Q^2$ or $m_f^2/M_\rho^2$ in the
electroproduction scattering
amplitude, where $Q^2$ is the photon virtuality. 
For transverse polarization, the two prescriptions lead to
a slightly different normalization but this can be attributed to the
ignoring of higher Fock
components in the wavefunction (since all of the normalization
integrals are in any case only unity up to corrections due to higher Fock components).

To compare with the HERA data we have to fix a value for the quark
mass. We take $m_f = 140$~MeV, which is the value used in the
fits to the deep-inelastic structure function, $F_2(x,Q^2)$ \cite{Forshaw:2011yj}.
We use the CGC[0.74] dipole model \cite{Soyez:2007kg,Watt:2007nr}  (see \cite{Forshaw:2011yj} for details),
although the predicitions do not
vary much if we use other models that fit the HERA $F_2$ data
\cite{Forshaw:2004vv,Kowalski:2006hc}. 
We also take $\kappa=M_\rho/\sqrt{2}=0.55$~GeV, which is the AdS/QCD prediction. 

\begin{figure}
\centering
\subfigure[~H1]{\includegraphics[width=0.85\textwidth]{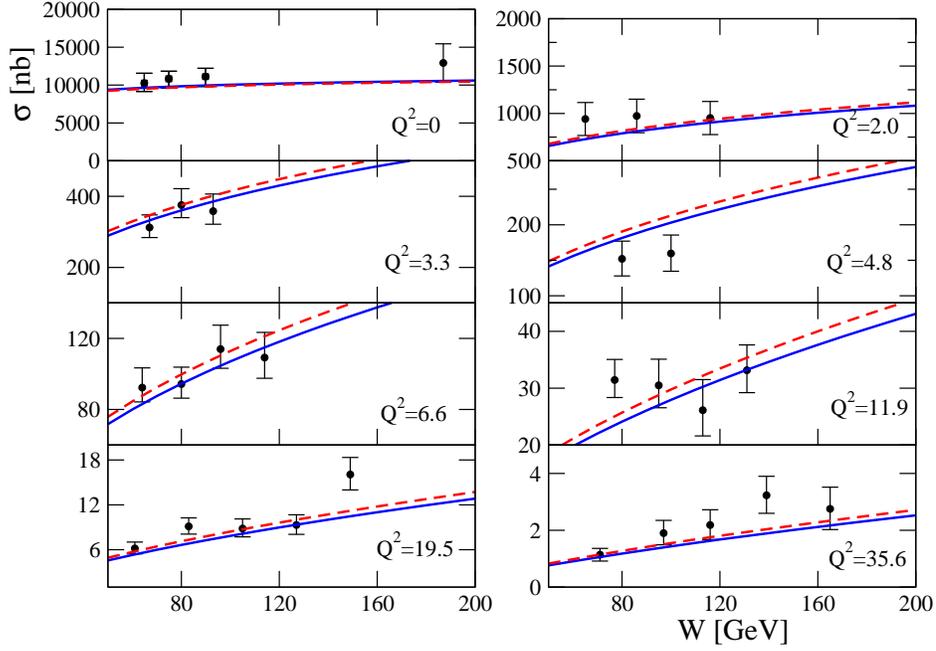} }
\subfigure[~ZEUS]{\includegraphics[width=0.85\textwidth]{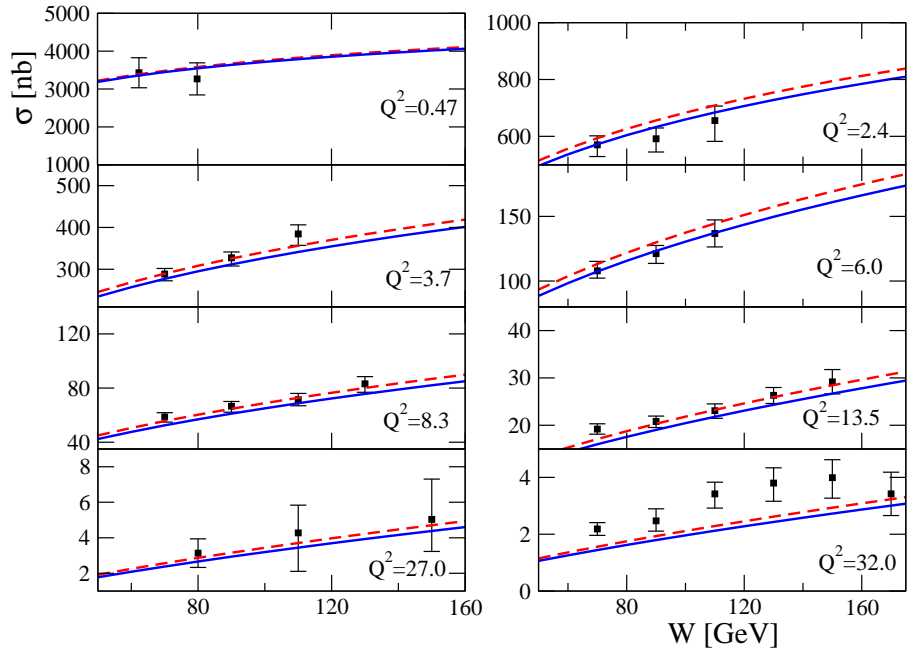} }
\caption{Comparison the HERA cross-section data
  \cite{Chekanov:2007zr,Collaboration:2009xp}. Solid blue curve is the
  AdS/QCD prediction and the dashed red curve is the best fit.}
\label{fig:xsecW}
\end{figure}

\begin{figure}
\centering
\includegraphics[width=0.7\textwidth]{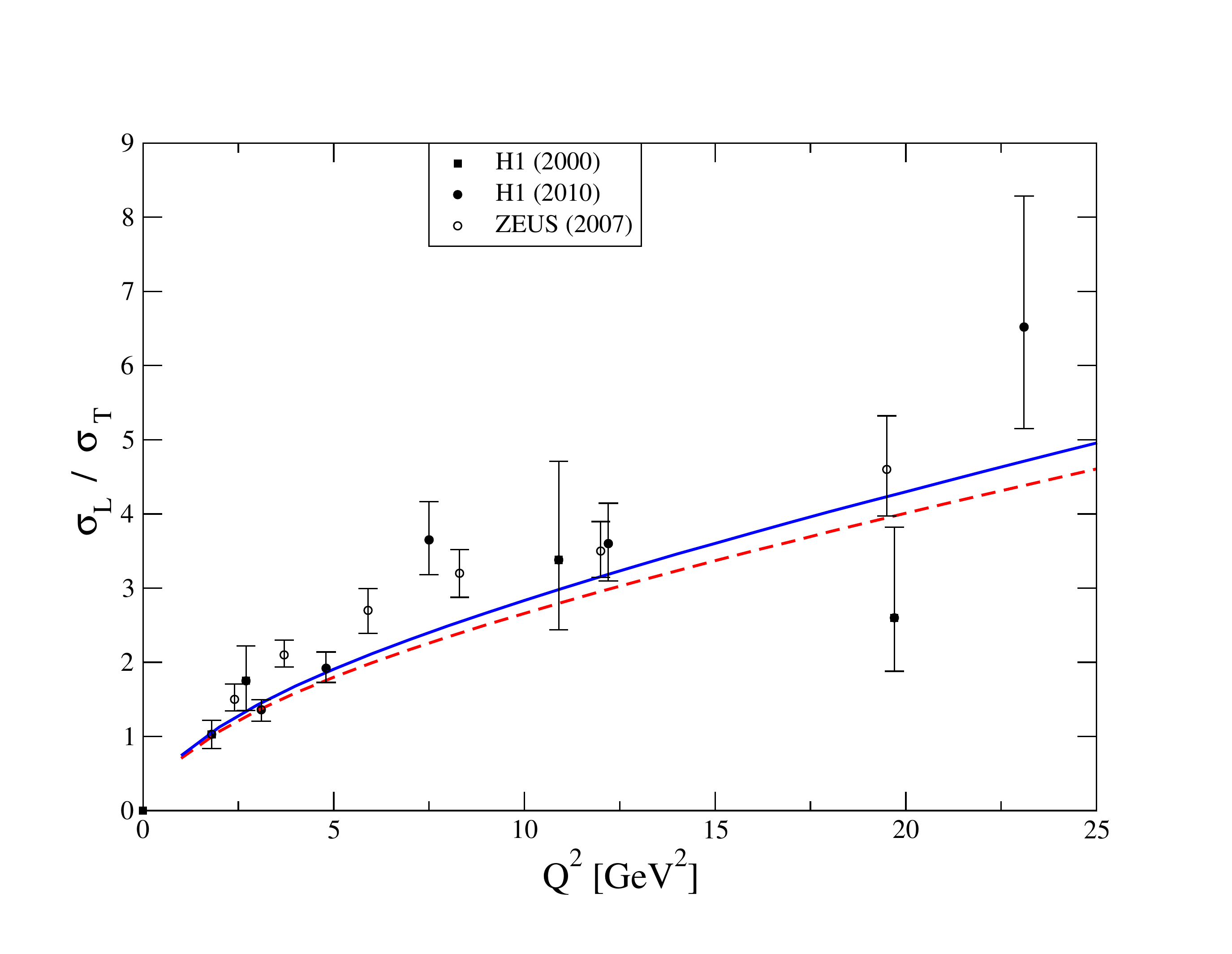}
\caption{Comparison to the HERA data on the longitudinal to transverse cross-section ratio
  \cite{Chekanov:2007zr,Collaboration:2009xp}. Solid blue curve is
 the AdS/QCD prediction and the dashed red curve is the best fit.}
\label{fig:ratio}
\end{figure}

Figures~\ref{fig:xsecW} and \ref{fig:ratio} compare the AdS/QCD predictions, shown as the solid blue curves, with the HERA
data. The agreement is
good and the disagreement at high $Q^2$ is not
unexpected. This is the region where perturbative evolution of the
wavefunction will be relevant and the AdS/QCD wavefunction we use is
clearly not able to describe that. It should be stressed that these AdS/QCD
predictions are parameter-free. 

We are also able to compute the electronic decay width
$\Gamma_{e^+e^-}$, which is related to the decay constant via
$$ f_\rho  = \left(  \frac{3\Gamma_{e^{+}e^{-}} M_\rho}{4 \pi
    \alpha_{\mathrm{em}}^2} \right)^{1/2}.$$
Using 
\begin{equation}
 f_\rho  = \frac{1}{2}
\left(\frac{N_c}{\pi}\right)^{1/2} \int_0^1 \d x \left(1 + \frac{m_{f}^{2}-\nabla^{2}}{M_\rho^2x(1-x)}\right) \phi_L(x,\zeta=0)~, 
\label{longdecayB}
\end{equation}
we obtain $\Gamma_{e^{+}e^{-}}=6.66$~keV, which is to be compared with the measured value $\Gamma_{e^{+}e^{-}}=7.04 \pm 0.06~\mathrm{keV}$
\cite{Nakamura:2010zzi}. 

\begin{figure}
\includegraphics[width=0.8\textwidth]{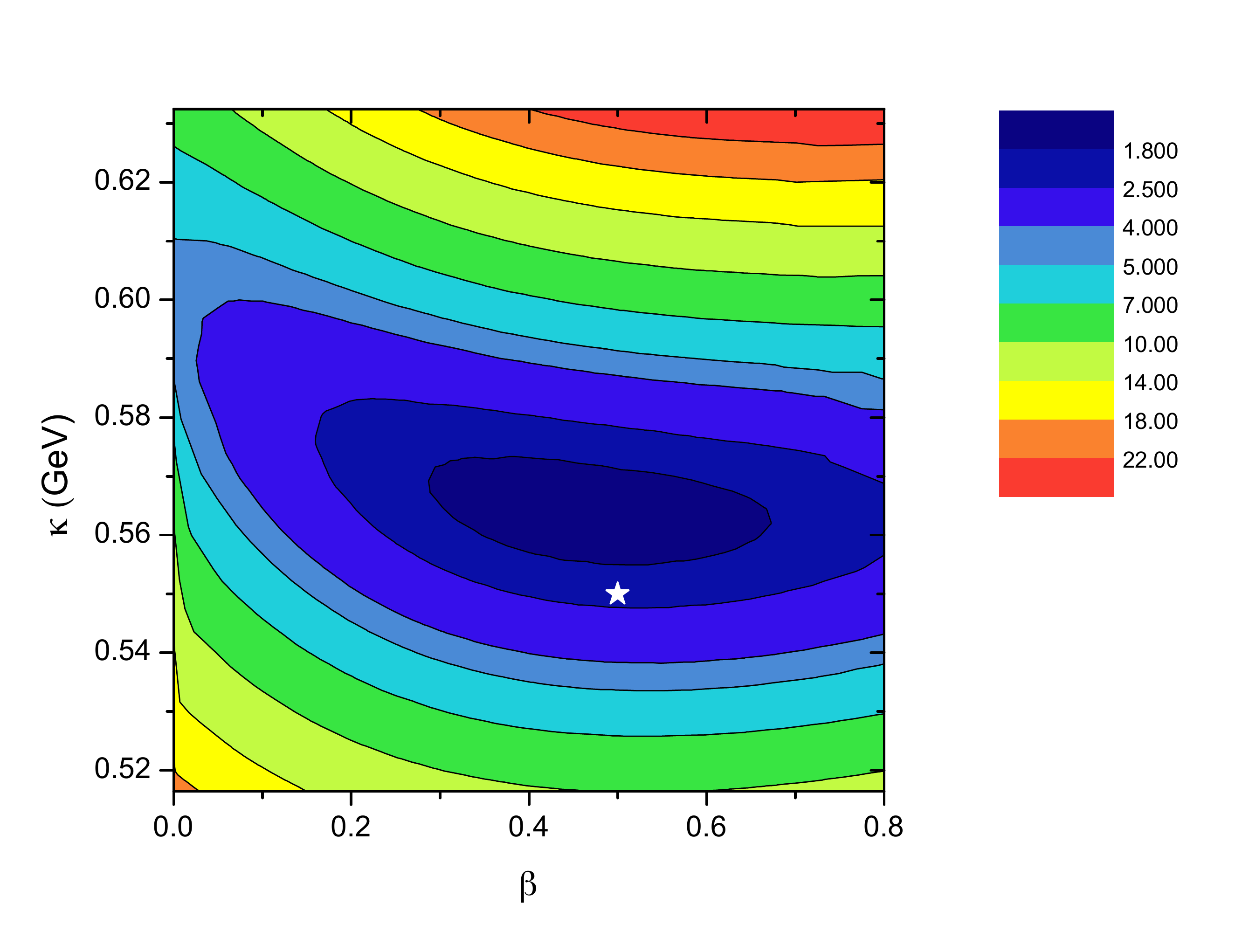}
\caption{The $\chi^2$ distribution in the $(\beta,\kappa)$ parameter space.  The AdS/QCD prediction is the white star.}
\label{fig:contour}
\end{figure}

Figure \ref{fig:contour} shows the $\chi^2$ per data point  in the
$(\beta,\kappa)$ parameter space (see
Eq. (\ref{lcwf-massive-quarks-fit})) \footnote{We include the
  electroproduction data and also the decay constant $f_\rho$ in the fit.}. It confirms that the AdS/QCD
prediction lies impressively close to the minimum in $\chi^2$. 
The best fit has a $\chi^2$ per data point equal to $114/76$ and is
achieved with $\kappa=0.56$ GeV and $\beta=0.47$. Better fits to the data are
possible, e.g. if one allows the parameters $\beta$ and $\kappa$ to
depend on the polarization of the meson.  However, given that we have
not attempted to quantify the theory uncertainty we regard these as
good fits.  The 
curves resulting from the best fit are shown as the red dashed curves
in Figures~\ref{fig:xsecW} and \ref{fig:ratio}.

We have previously shown that the twist-$2$
Distribution Amplitude (DA) can be related to $\phi_L(x,\zeta)$
according to 
\begin{equation}
\varphi(x,\mu) = \left(\frac{N_c}{\pi}\right)^{1/2}  \frac{1}{2f_\rho}
\int \d
b \; \mu
J_1(\mu b) \left(1  + \frac{m_f^2 -\nabla^2}{M_\rho^2x(1-x)} \right) \phi_L(x,\zeta)~.
\label{tw2DAB}
\end{equation}
We note that $\int \d x \; \varphi(x,\mu \to \infty)=1$ recovers the decay
constant constraint.  To compare to predictions
using QCD Sum Rules \cite{Ball:2007zt} and from the lattice \cite{Boyle:2008nj},  we
can also compute the moment:
\begin{equation}
 \int_0^1 \d x \;  (2x-1)^2 \varphi(x,\mu)~.
\end{equation}
We obtain a value of $0.228$ for the AdS/QCD wavefunction, which is to be compared with the
Sum Rule result of $0.24 \pm 0.02$ at $\mu = 3$~GeV \cite{Ball:2007zt} and the lattice
result of $0.24\pm 0.04$ at $\mu = 2$~GeV \cite{Boyle:2008nj}.  The AdS/QCD
wavefunction neglects the perturbatively known evolution with the
scale $\mu$ and should be viewed as a parametrization of the DA at
some low scale $\mu \sim 1$ GeV. Viewed as such, the agreement is
good. 

\section{Acknowledgements}
We thank Stan Brodsky, Mike Seymour and Guy de T\'eramond for their
helpful comments and suggestions.
The work of JRF is supported by the Lancaster-Manchester-Sheffield
Consortium for Fundamental Physics under STFC grant ST/J000418/1. R.S thanks the University of Manchester and the Institute for Nuclear Theory at the University of Washington for hospitality and financial support.

\bibliography{letter}

\end{document}